\documentclass[twocolumn,showpacs,preprintnumbers,amsmath,amssymb]{revtex4}
\usepackage{epsfig}

\usepackage{graphicx}
\usepackage{dcolumn}
\usepackage{bm}

\newcommand{\be}{\begin{equation}}
\newcommand{\ee}{\end{equation}}
\newcommand{\bea}{\begin{eqnarray}}
\newcommand{\eea}{\end{eqnarray}}
\newcommand{\beq}{\begin{equation}}
\newcommand{\eeq}{\end{equation}}
\newcommand{\nn}{\nonumber}

\begin{document}

\title{Proton-proton diffractive collisions
 at ultra-high energies}

\author{V.V. Anisovich}
\affiliation{National Research Centre "Kurchatov Institute",
Petersburg Nuclear Physics Institute, Gatchina, Russia}
\author{K.V. Nikonov}
\affiliation{National Research Centre "Kurchatov Institute",
Petersburg Nuclear Physics Institute, Gatchina, Russia}
\author{V.A. Nikonov}
\affiliation{National Research Centre "Kurchatov Institute",
Petersburg Nuclear Physics Institute, Gatchina, Russia}
\affiliation{Helmholtz-Institut f\"ur Strahlen- und Kernphysik,
Universit\"at Bonn, Germany}

\date{\today}

\begin{abstract}
Proton-proton total and elastic cross sections are considered in the
Dakhno-Nikonov eikonal model [L.G. Dakhno, V.A. Nikonov, Eur. Phys.
J. A{\bf 8}, 209 (1999)] at ultra-high energies. The model takes
into account the quark structure of hadrons and the gluon structure
of the supercritical pomeron that results in colour screening. The
pomeron is considered as an input interaction term. The model gives
a reasonably good description of the preLHC and LHC data for $pp$
collisions with a growth of the type $\ln^2s$ for  total and elastic
cross sections and $(\tau={\bf q}_\perp^2\ln^2s)$-scaling for
diffractive scattering. We present parameters of the supercritical
pomeron and provide predictions for the energy region $\sqrt{s}\sim
10^2-10^4 $ TeV.
\end{abstract}
\maketitle

\section{Introduction}

The observed growth of total cross sections at preLHC energies
\cite{pre} initiates studies of corresponding models such as that
with a maximal increase allowed by the Froissart bound \cite{Froi}
or with "heretical" power-$s$ behavior \cite{Kaid,Land}. The
$s$-channel unitarization of scattering amplitudes actualizes the
use of the Glauber approach. On account of the $s$-channel
rescatterings the power-$s$ growth of amplitudes is dampened to the
$(\ln^2 s)$-type, see, for example, \cite{Gaisser,Block,Fletcher}.

The eikonal method for the $s$-channel high-energy unitarization of
the scattering amplitude was used in the Dakhno-Nikonov model
\cite{DN}. The model takes into account the quark structure of
colliding hadrons, the gluon origin of the input pomeron and colour
screening effects in collisions. The model provided a satisfactory
description of $\pi p$, $p p$, $\bar p p$ data for diffractive
processes at preLHC energies, $\sqrt s \sim 0.2-1.8$ TeV.

Recent measurements at LHC (ATLAS, CMS, TOTEM collaborations) and
cosmic ray data initiate further interest to $s$-channel unitarized
amplitudes, see, for example,
\cite{1110.1479,1111.4984,1201.6298,1208.4086, 1212.5096}
and references therein.

In the present note we consider diffractive $pp$ scatterings in
terms of the Dakhno-Nikonov model, concentrating our attention on
the ultra-high energy asymptotic behavior. We refit data taking into
account new ones in the TeV-region \cite{totem,auger}. According to
the fit, this region is a pre-asymptotic one. The asymptotic
behaviour starts, actually, at $10^2-10^3$ TeV.

For the ultra-high energy limit the Dakhno-Nikonov model gives for
total and elastic $pp$ cross sections a $(\ln^2s)$-growth:
$\sigma_{tot}\sim \ln^2s$ and $\sigma_{el}\sim\ln^2s$. The high
energy cross sections ($\sigma_{el}$, $\sigma_{tot}$) are
approaching their asymptotic values from bottom to top:
$\sigma_{tot}(s)/\sigma_{tot}^{(asym)}(s)< 1 $; this gives the
illusion of exceeding the Froissart bound (though, let us emphasize,
exceeding the Froissart bound does not violate general constraints
for scattering amplitudes \cite{azimov}).

The model tells that differential elastic cross sections depend
asymptotically on transverse momenta with realization of
$\tau$-scaling $(\tau={\bf q}_\perp^2\ln^2s)$:
 \be \label{3}
\frac{d\sigma_{el}(\tau)}{d\tau}=
  D(\tau ),\quad {\rm with}\quad
\int\limits_0^\infty d\tau D(\tau)=\sigma_{el}(s)\sim \ln^2s   .
\ee
 Formulae of the Dakhno-Nikonov model which are used for the calculation of
$\sigma_{tot}$, $\sigma_{el}$, $d\sigma_{el}/dq^2_\perp$ are given
in the next section. In section 3 we present results of the fit and
predictions for ($10^2-10^4$)-TeV region.

\section{ Formulae for diffractive hadron-hadron scattering}

\begin{figure*}
\centerline{\epsfig{file=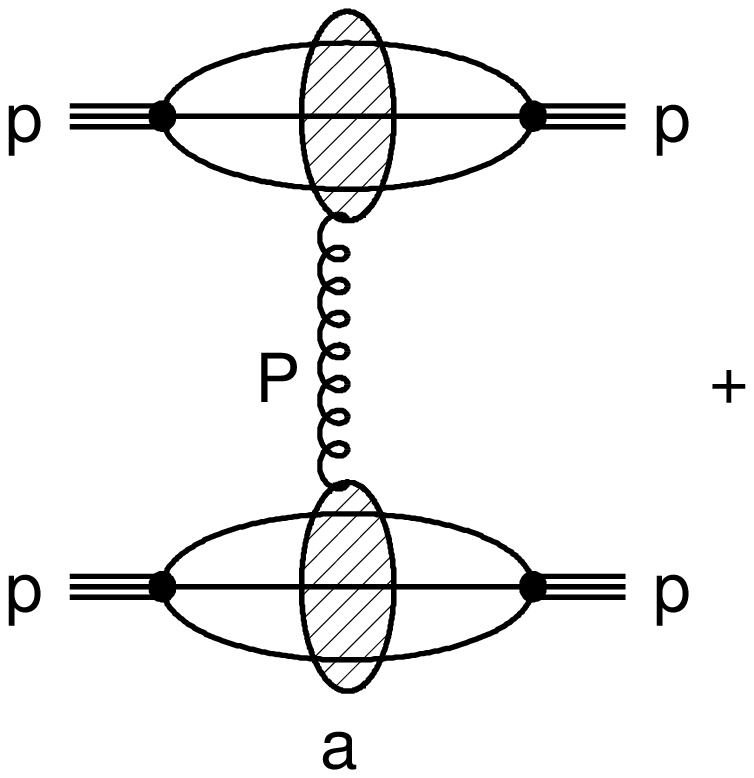,width=3cm}\hspace{0.2cm}
            \epsfig{file=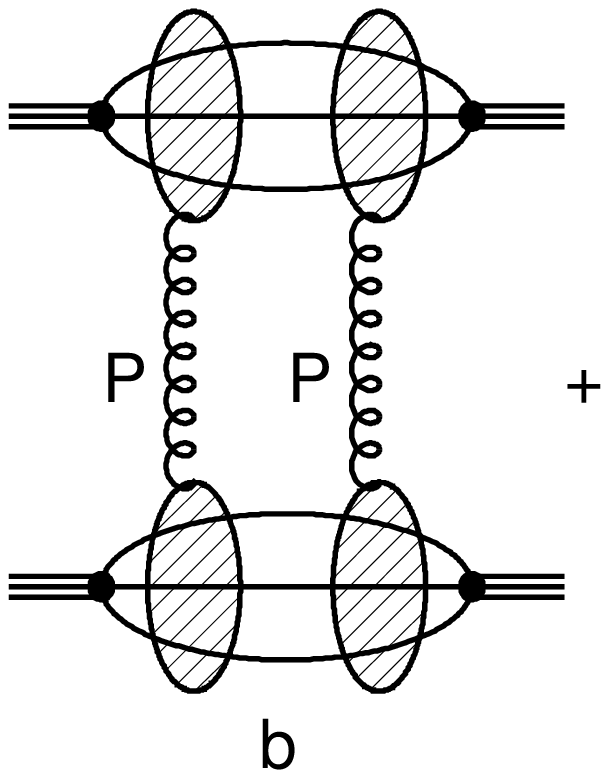,width=3cm}\hspace{0.2cm}
            \epsfig{file=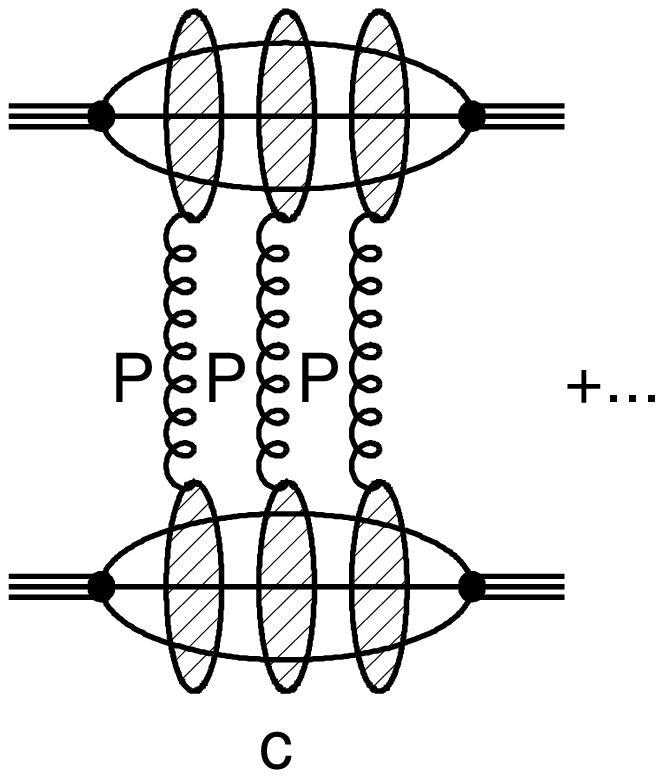,width=3cm}}
\centerline{\epsfig{file=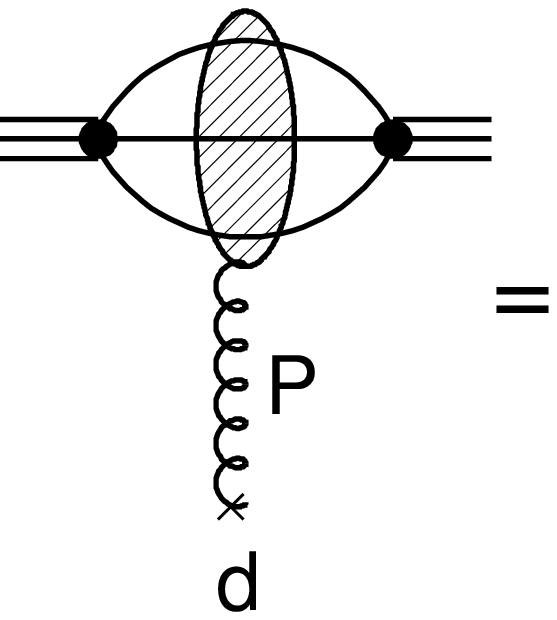,height=3cm}
            \epsfig{file=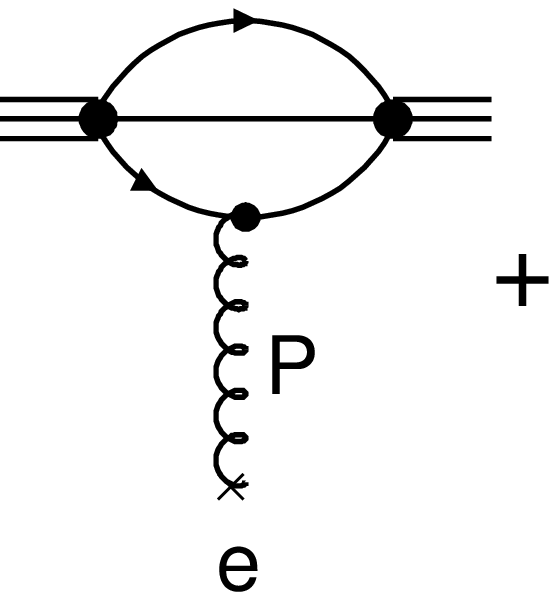,height=3cm}
            \epsfig{file=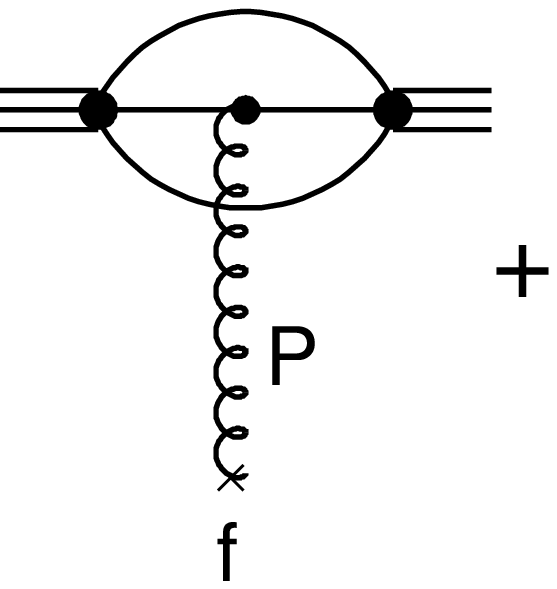,height=3cm}
            \epsfig{file=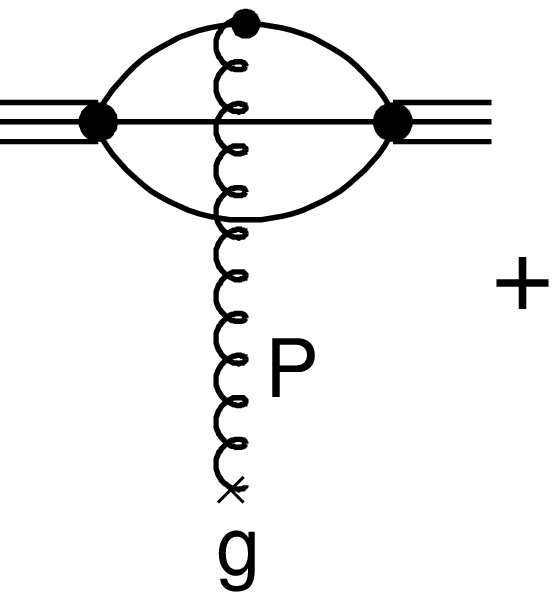,height=3cm}
            \epsfig{file=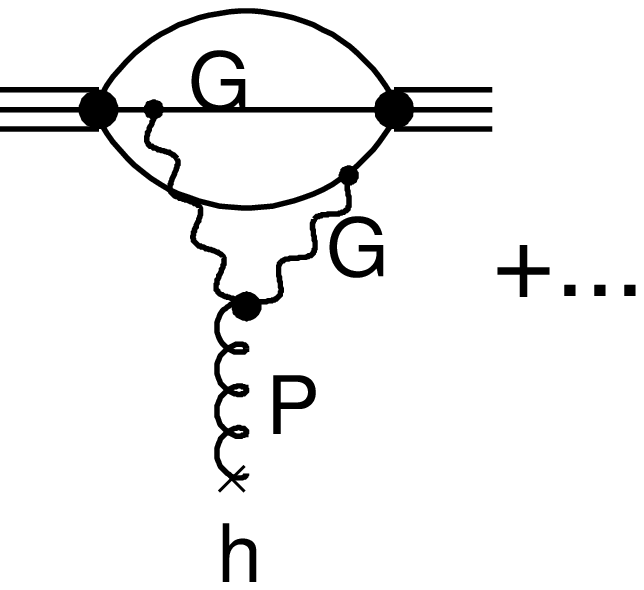,height=3cm}}
\caption{Proton-proton collisions: (a)--(c) diagrammatic
representation of the $p p$ scattering
amplitude as a set of $s$-channel pomeron ($P$) interactions;
(d)--(h) pomeron-proton vertex and its deciphering: three
pomeron-quark terms and three $GGP$ terms describe interaction of
gluons ($G$) of pomeron with quarks of proton. \label{f1} }
\end{figure*}

The model is based on the hypothesis of the gluon origin of the
$t$-channel forces, and these gluons form pomerons. Hadrons, mesons
(two-quark composite systems) and baryons (three-quark composite
systems), scatter on the pomeron cloud. It is supposed that the
pomeron cloud is materialized as a low-density gas, and
pomeron-pomeron interactions, as well as $t$-channel transitions
$P\to PP$, $P\to PPP$ and so on, can be neglected. Consequently, the
$pp$ scattering amplitude is determined by the set of diagrams shown
in Fig. \ref{f1}.

The pomerons are formed by effective gluons ($G$) which are massive,
$\sim 700-1000$ MeV \cite{parisi,field}. The pomeron parameter
$\alpha'_P$ is small $\alpha'_P\simeq(0.10-0.25)$ GeV$^{-2}$, that
means pomerons are comparatively heavy and hard \cite{gribov-hard}.
The gluon structure of the pomeron provides colour screening effects
for hadron quarks \cite{adnPR}.

 Total and elastic cross sections
of the model for colliding hadrons $A$ and $B$ are
written as follows:
\bea \label{4}
\sigma_{tot}(AB)&=&2\int d^2b
\int dr'\varphi^2_A(r')dr''
\varphi^2_B(r'')
\nn \\
&\times&
\left [1-\exp{(-\frac{1}{2}
\chi_{AB}(r',r'',{\bf b})})\right],  \nn \\
\sigma_{el}(AB)&=&\int d^2b
 \bigg(\int dr'\varphi^2_A(r')dr''\varphi^2_B(r'')
\nn \\
&\times&
\left[1-\exp{(-\frac{1}{2}\chi_{AB}(r',r'',{\bf b})})\right]
\bigg)^2 ,
\nn \\
4\pi\frac{d\sigma_{el}}{dq^2_\perp}(AB)&=&
\int d^2b\ e^{i{\bf q_\perp} {\bf b}}\int dr'\varphi^2_A(r')dr''
\varphi^2_B(r'')
\nn \\
&\times&
\left[1-\exp{(-\frac{1}{2}\chi_{AB}(r',r'',{\bf b})})\right]
\nn \\ &\times&
\int d^2\tilde b\ e^{-i{\bf q_\perp}{\bf \tilde b}}
 \int d\tilde r'\varphi^2_A(\tilde r')d\tilde r''\varphi^2_B(\tilde r'')
\nn \\
&\times&
\left[1-\exp{(-\frac{1}{2}\chi_{AB}(\tilde r',\tilde r'',
{\bf \tilde b})})\right].
\eea
Here $dr\varphi^2_{A}(r)$, $dr\varphi_{B}^2(r)$ are the quark densities
of  colliding hadrons:
\bea \label{5}
dr \varphi^2_{\pi}(r)&=&d^2r_1d^2r_2\delta ^{(2)}( {\bf r}_1+{\bf r}_2)
\varphi^2_{\pi}(r_1,r_2),\nn \\
dr \varphi^2_p(r)&=&d^2r_1d^2r_2d^2r_3 \delta ^{(2)}( {\bf r}_1+{\bf
r}_2+{\bf r}_3) \varphi^2_p(r_1,r_2,r_3),\nn \\
\eea
 where ${\bf r}_a$ are the transverse coordinates of  quarks, and
$\varphi_{A}^2$, $\varphi_{B}^2$ are given by quark wave functions
squared integrated over longitudinal variables. Proton and pion
quark densities are determined using the corresponding form factors;
such an estimation can be found, for example, in \cite{AMN}. The
profile-function $\chi_{AB}$ describes the interaction of quarks via
pomeron exchange as follows:
\bea
\label{6}
\chi_{AB}(r',r'',{\bf b})&=&\int d^2b'd^2b''\delta^{(2)} ({\bf b}-{\bf b'}+{\bf
b''})
\nn \\
&\times&
S_A(r',{\bf b'})S_B(r'',{\bf b''}).
\eea
Functions $S_{A,B}$ stand for the pomeron-quark interactions; they are
determined by the diagrams with different couplings of the pomeron
quarks:
\bea
\label{7}
S_{\pi}(r,{\bf b})&=&\rho( {\bf b}- {\bf
r}_1)+\rho( {\bf b}- {\bf r}_2)
\nn \\
&-& 2\rho( {\bf b}-\frac{ {\bf r}_1+
{\bf r}_2}{2}) \exp(-\frac{( {\bf r}_1-
 {\bf r}_2)^2}{4r^2_{cs}}), \nn \\
S_p( {\bf r}, {\bf b})&=&
\sum\limits_{i=1,2,3}\,\rho( {\bf b}- {\bf r}_i)
\nn \\
&-&\sum\limits_{i\ne k}\,
\rho( {\bf b }-\frac{ {\bf r}_i+ {\bf r}_k}{2})
\exp(-\frac{( {\bf r}_i-
 {\bf r}_k)^2}{4r^2_{cs}}).
\eea The term $\rho( {\bf b}- {\bf r}_i)$ describes the diagram
where the pomeron couples to one of the hadron quarks while the
terms proportional to $\exp(-r^2_{ij}/r^2_{cs})$ are related to the
diagram with the pomeron couples to two quarks of the hadron. Such a
diagram is a three-reggeon graph $GGP$ where $G$ is the reggeized
gluon. Functions $S_{\pi}$ and $S_p$ tend to zero as $|\vec
{r}_{ij}| \to 0$: this is the colour screening phenomenon inherent
to gluonic pomeron. For the sake of convenience, one can perform
calculations in the center-mass-system of the colliding quarks,
supposing that the hadron momentum is shared equally between its
quarks. Then
\beq
\label{8}
\rho({\bf b})= \frac
{g}{4\pi(G+\frac{1}{2}\alpha'_P\ln{s})}
 \exp\left [-\frac{{\bf b}^2}{4(G+\frac{1}{2}\alpha'_P\ln{s})}\right],
\eeq
where the vertex $g$ depends on the energy squared
of the colliding quarks, $s_{qq}$:
\beq
g^2=g_0^2+g_1^2\left ( \frac{s_{qq}}{s_0} \right )^{\Delta}     .
\eeq
Below
$s_0=1$ GeV$^2$.
 The parametrization of $g^2$ corresponds to the two-pole
presentation of the QCD-motivated pomeron with
intercepts $\alpha(0)=1$ and $\alpha(0)=1+\Delta$.

Let us consider $pp$ scattering, specifically, the term with $qq$
interaction, see Eqs. (\ref{6}), (\ref{8}):
\bea
\label{10}
&&
\chi^{(q_i q_j)}_{pp}(r',r'',{\bf b})=\int d^2b'd^2b''\delta^{(2)}
({\bf b}-{\bf b'}+{\bf b''}) \nn \\
&& \times
 \frac {g}{\pi(4G+2\alpha'_P\ln{s})}
\exp\left [-\frac{({\bf b'}-{\bf r'}_i)^2}{4G+2
\alpha'_P\ln{s}}\right]\nn \\
&& \times
 \frac {g}{\pi(4G+2\alpha'_P\ln{s})}
\exp\left [-\frac{({\bf b'}-{\bf r''}_j)^2}{4G+2
\alpha'_P\ln{s})}\right]\nn \\
&&
=
\frac{g^2}{2\pi(4G+2\alpha'_P\ln{s})}
\exp{\Big[-\frac{({\bf b}-{\bf r'}_i+{\bf r''}_j)^2}{2 (4G+2\alpha'_P\ln{s})}\Big]}
. \quad
\eea
Equations depend on the transverse coordinates of quarks,
 though the original expressions depend on the fractions of the
momenta of the colliding hadrons carried by the quark, $x_i$. In the
functions $S_\pi$, $S_p$ we put $x_i=1/2$ for a meson and $x_i=1/3$
for the proton, in other words we assume that hadron wave functions
$\varphi_{A}(r,x)$ and $\varphi_B(r,x)$ select the mean values of
$x_i$ in the interaction blocks. So, in the used equations we put
$s_{qq}=s/9$ for $p p$ collisions and perform the renormalization of
the value  $G$ by including the factor $\alpha'_P\ln{9}$ that can
lead to $G<0$.
Of course, the slope parameter $2G+\alpha'_P\ln{s}$ should be
positive in the region of application of the model.

\begin{figure*}[t]
\centerline{\epsfig{file=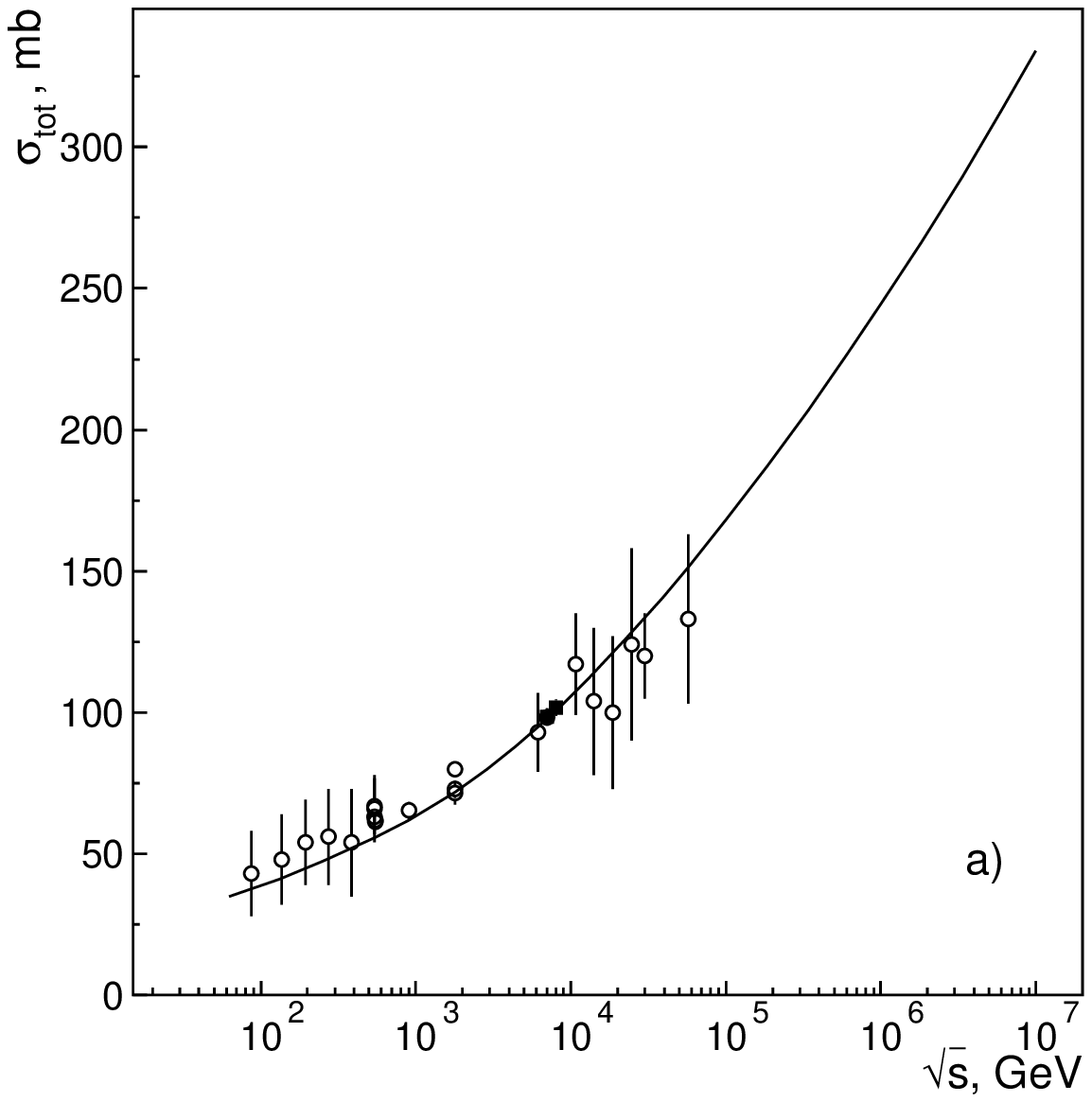,width=7cm}
            \epsfig{file=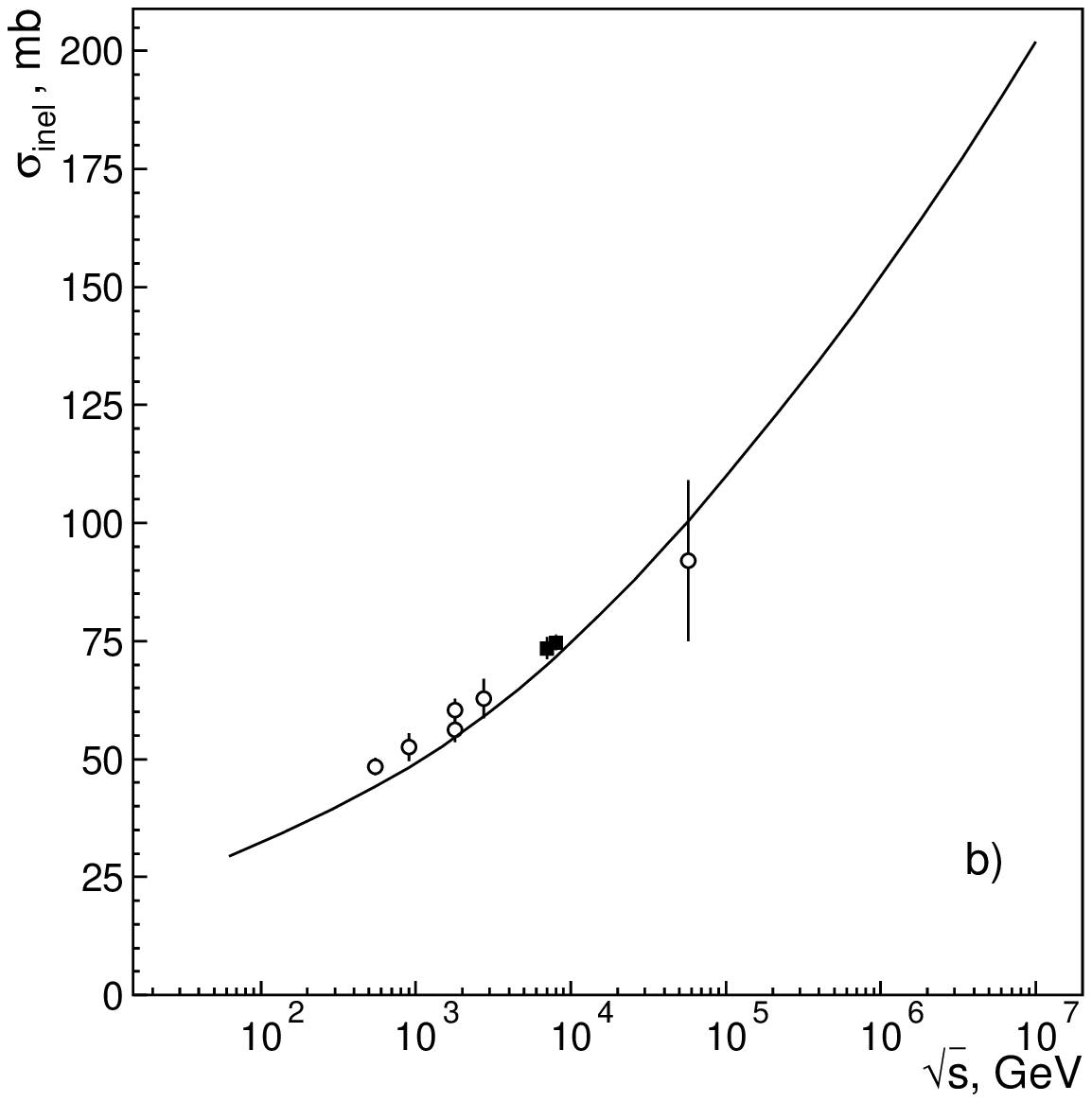,width=7cm}}
\caption{a) Total and b) inelastic cross section data
\cite{pre,totem,auger} and fit in the Dakhno-Nikonov model.} \label{f2}
\end{figure*}

\begin{figure*}[ht]
\centerline{\epsfig{file=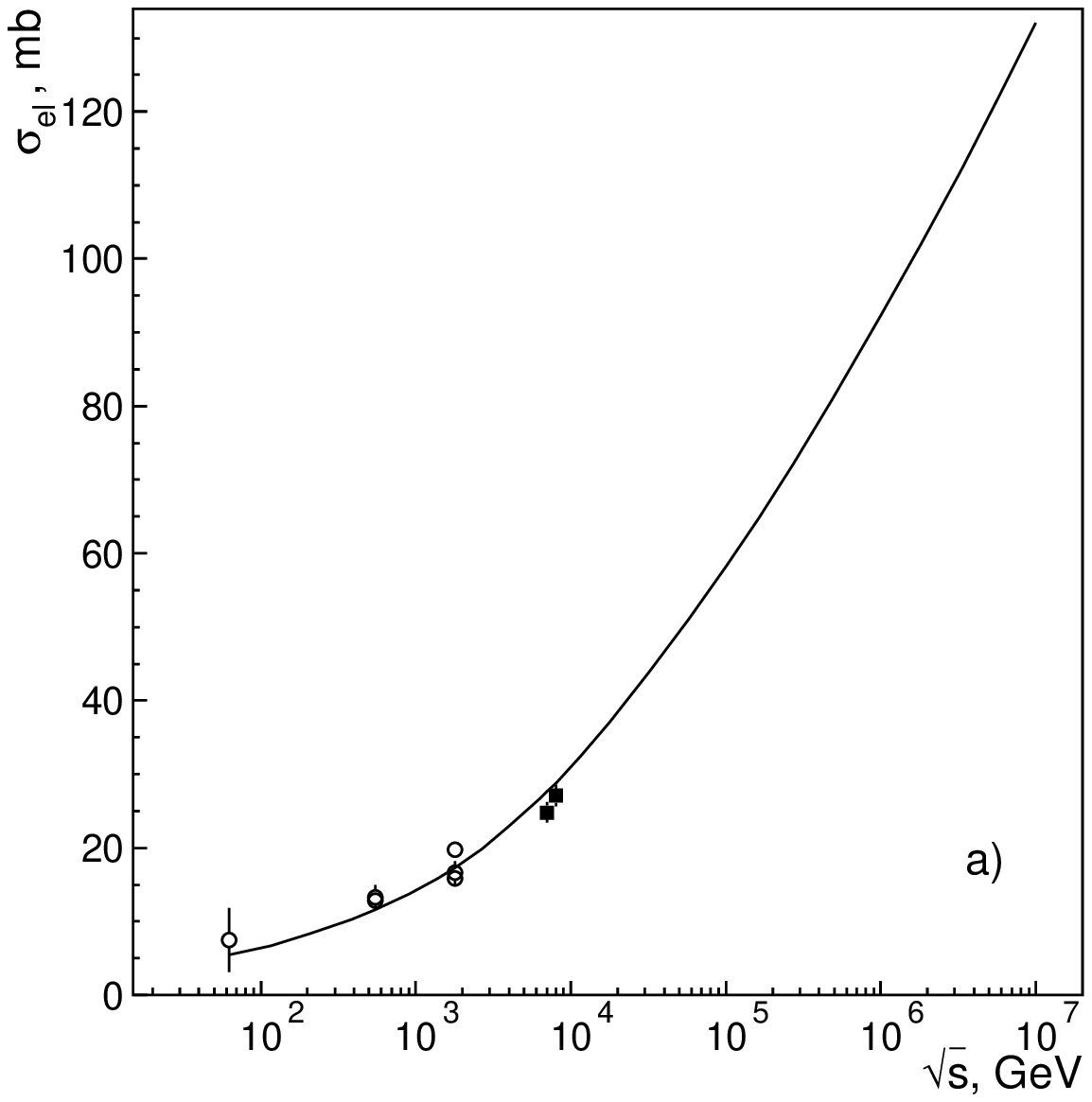,width=7cm}
            \epsfig{file=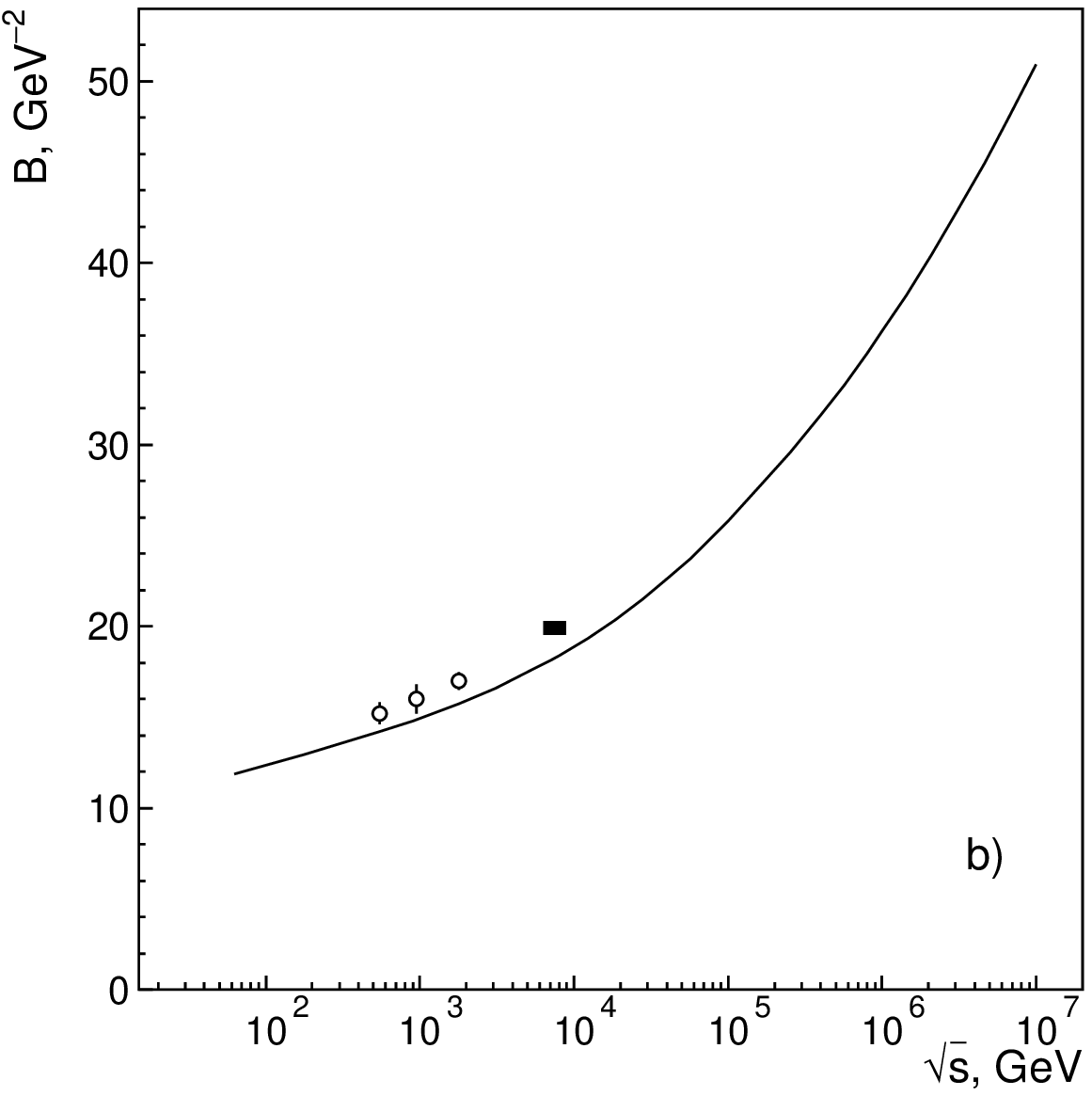,width=7cm}}
\caption{a) Elastic cross section  and b) diffractive cone slope $B$
determined as $d\sigma_{el}/d{\bf q}^2_\perp
=B\sigma_{el}\exp(-B{\bf q}^2_\perp)$ . Results of the fit are given
by solid curves. }
\label{f3}
\end{figure*}

The Dakhno-Nikonov model is actually a realization of the
Good-Walker eikonal approach \cite{GW} for a continual set of
channels (each quark configuration with fixed coordinates is a
separate channel). Used in the model the two-pole pomeron exchange
is popular from the sixties till now, see for example ref. \cite{DL}.

\section{Fit of the high energy $pp$ data
and predictions for the $(10^2-10^4$)-TeV region}

\begin{figure*}[ht]
\centerline{\epsfig{file=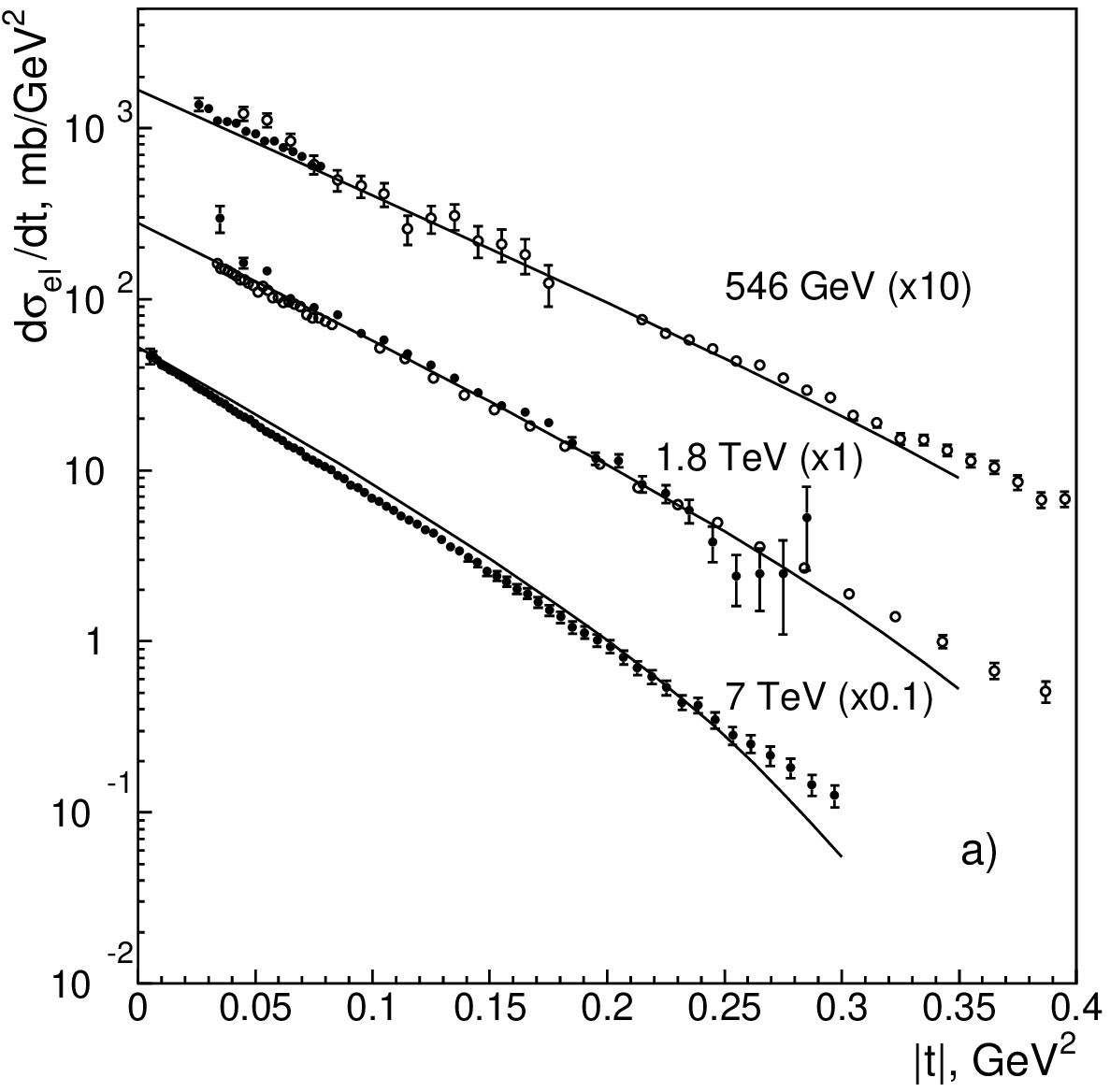,width=8cm}\hspace{0.3cm}
            \epsfig{file=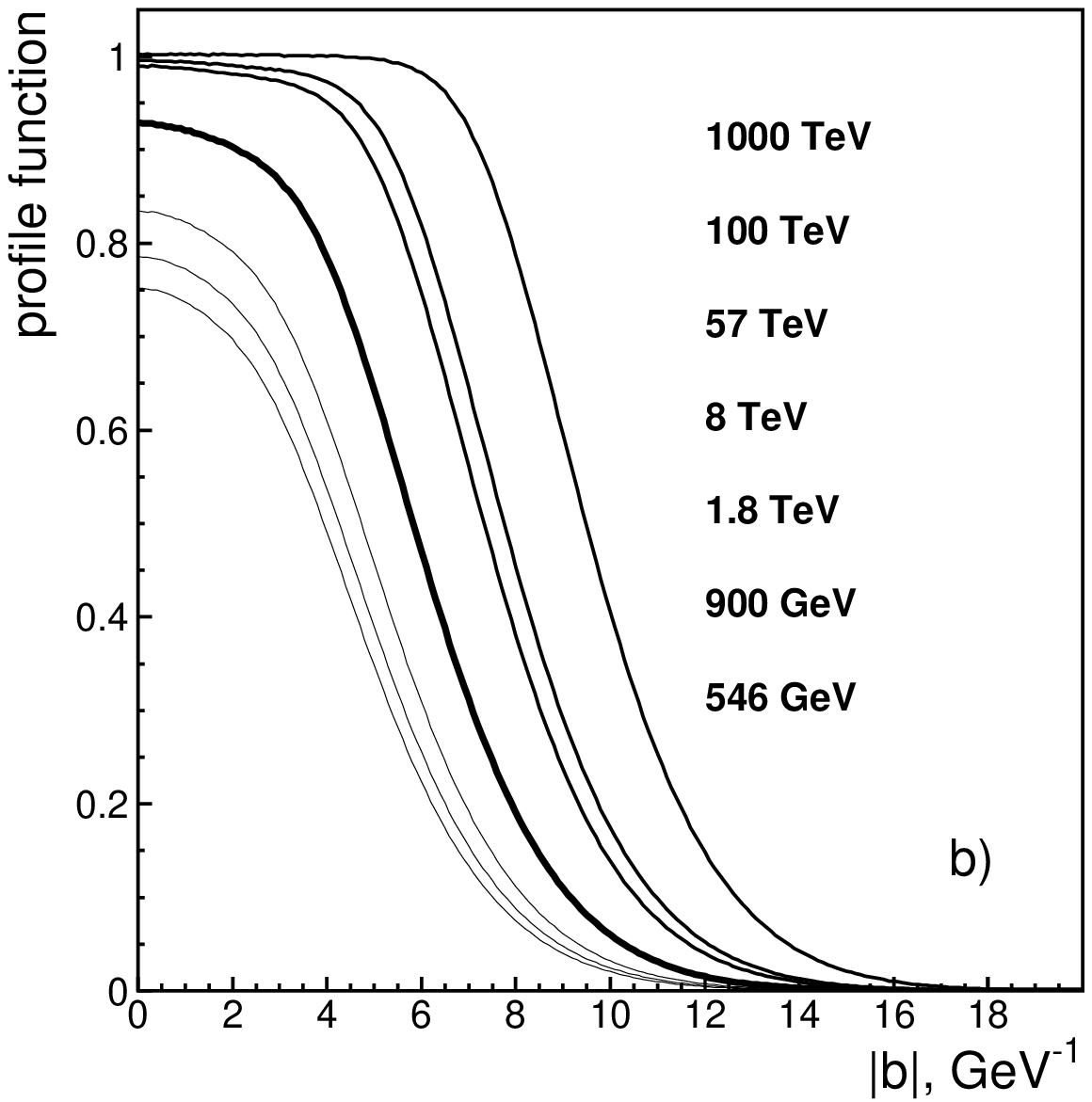,width=8cm}}
\caption{ a) Differential cross section $ d\sigma_{el}/d{\bf
q}^2_\perp$ at $\sqrt s =0.546, 1.8, 7.0$ TeV. b) Profile functions
$T(b)$ determined in Eq.
(\ref{p9})
 at preLHC (0.546-1.8 TeV), LHC (8.0 TeV) and ultra-high
(100-1000 TeV) energies .}
\label{f4}
\end{figure*}

Figs. \ref{f2}-\ref{f4} demonstrate fit results of the diffractive
scattering $pp$ data, including those at LHC energies \cite{totem}
and cosmic ray ones \cite{auger}. In Fig. \ref{f2} we show total and
inelastic cross sections, the fit gives a good approximation to data
at $\sqrt{s}\sim 50-5\cdot 10^4$ GeV and a prediction for the region
$\sqrt{s}\sim 10^5- 10^7$ GeV.
 The same level of description demonstrates us $\sigma_{el}$ and the
slope $B$, see Fig. \ref{f3}.

In Fig. \ref{f4}a we show
$d\sigma_{el}/d{\bf q}^2_\perp $ at ISR and LHC energies and their
fit, and
Fig. \ref{f4}b demonstrates the profile function $T(b)$ determined as
\bea \label{p9}
&&
\sigma_{tot}=2\int d^2b\; T(b)=2\int d^2b
\Big[1-e^{-\frac12\chi(b)}\Big], \nn \\
&&
4\pi\frac{d\sigma_{el}}{d{\bf q}^2_\perp}=
A^2({\bf q}^2_\perp),\quad
A({\bf q}_\perp)=\int d^2b e^{i{\bf b}{\bf q}_\perp} T(b).
\quad
\eea

%
%
%

Obtained in the fit the profile functions $T(b)$, which are given by
Eq. (\ref{p9}), are shown in Fig. \ref{f4}b for preLHC, LHC and
ultra-high energies. It is seen that the profile function saturation
mode, $T(b)\to 1$, works at ultra-high energies ($100-1000$ TeV), at
LHC energies the saturation mode is only starting.

Parameters of the present fit and that of ref. \cite{DN}
approximately coincide:
\be
\begin{tabular}{l|l|l}
      & fit\,of\, ref.\, \cite{DN}& this \, fit \\
\hline
$\Delta$                    &  0.29  &  0.273 \\
$g^2_0$ [mb]                &  8.079 &  8.106 \\
$g^2_1$ [mb/GeV$^{2\Delta}$]  &  0.338 &  0.379 \\
$\alpha'_P$ [(GeV/c)$^{-2}$]&  0.25  &  0.129 \\
$G$ [(GeV/c)$^{-2}$]        & -0.40   & -0.365 \\
$r^2_{cs}$ [(GeV/c)$^{-2}$] &  0.80   &  0.67  \\
\end{tabular}
\ee
 The only change is in the value of $\alpha'_P$, we observe the
decrease of $\alpha'_P$ that means the slowing of the asymptotic
regime switch-on.


 At $\ln s>>1$, when the asymptotic regime works, there are well-seen
 two regions in the $b$-space (Fig. \ref{f4}b):
 with $T(b)\simeq 1$ (black disk) and $T(b)\simeq 0$ (transparent
 space).
 Conventionally we determine these areas by constraints:
\bea  \label{11}
&&
{\bf b}^2 < 4\Delta\alpha'_P\ln^2{\frac {s}{s_-}}\,,\qquad
{\rm with}\quad T(b)>0.97,
\nn \\
&& {\bf b}^2 > 4\Delta\alpha'_P\ln^2{\frac {s}{s_+}}\,,\qquad {\rm
with}\quad T(b)< 0.03
\quad
\eea
 The black disk area with radius
\be \label{12}
R_{black}=2\sqrt{ \Delta\alpha'_P}\,\ln{\frac
{s}{s_0}}\,,\qquad \sqrt{s_0}\simeq 80 {\rm GeV}
\ee
 reveals itself in the region of energies with the growth of the cross
sections.

 The black disk radius depends on parameters of the leading pomeron
 (factor $\Delta\alpha'_P$) that
is realized in the Gribov's equality of hadronic total cross
sections \cite{Gribov-tot} at asymptotic energies:
 $\sigma_{tot}(\pi p)/\sigma_{tot}(p
p)\to 1$, $\sigma_{tot}(\pi \pi)/\sigma_{tot}(p p)\to 1$, and so on.

\begin{figure}
\centerline{\epsfig{file=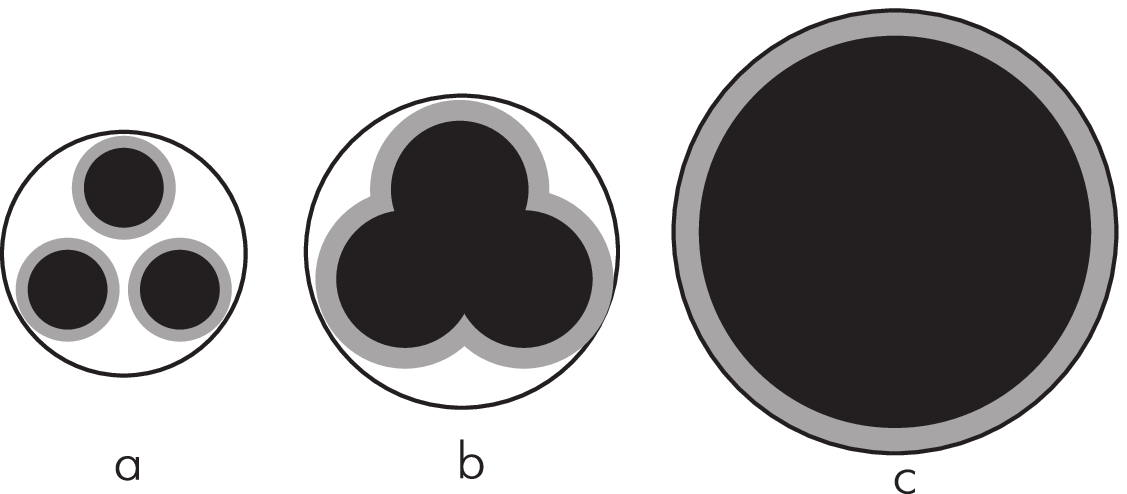,width=8cm}}
\caption{Proton picture in the impact parameter space at moderately
high energies (a) and its transformation with increasing energy
(b) to
ultra-high region (c).
\label{disks} }
\end{figure}

We perform unitarization of scattering amplitude supposing it is
originated by conventional pomerons though other types of the input
pomerons are possible as well as non-pomeron short-range
contributions (for example, see \cite{1201.6298,KL,KN,GLM,AKL}). But
here we concentrate our attention to peripheral interactions and its
transformation with energy growth. Small deviations of the fitting
curves from data can be easily improved with use some kind of
short-range contributions.

%
%
%

\section{Conclusion}

The twofold structure of hadrons -- hadrons are built by constituent
quarks and the latter are formed by clouds of partons -- manifests
itself in hadron collisions. At moderately high energies colliding
protons reveal themselves in the impact parameter space as three
disks corresponding to three constituent quarks, Fig. \ref{disks}a.
 At ultra-high energies the situation is transformed to a one-disk
 picture, Fig. \ref{disks}c, and the energy of this transformation is that of
LHC. The radius of the black disk
at asymptotic energies is increasing as $\ln{s}$ that
provides a $\ln^2s$ growth of $\sigma_{tot}$, $\sigma_{el}$ with
$\sigma_{el}/\sigma_{tot}\to 1/2$ and a $\tau$-scaling for
diffractive cross sections.

\section{Acknowledgment}

We thank Ya.I. Azimov, A.K. Likhoded, J. Nyiri and M.G. Ryskin for
useful discussions. The work
was supported by RFBR, grant 13-02-00425.

   \end{document}